\documentclass{elsart}

\usepackage{graphicx}
\usepackage{epsfig}
\usepackage{bbm}
\usepackage{amsmath,amssymb}

\newcommand{\dd}{\mathrm{d}}
\newcommand{\ee}{\mathrm{e}}
\newcommand{\RR}{\mathbbm{R}}
\newcommand{\PP}{\mathbbm{P}}
\newcommand{\EE}{\mathbbm{E}}

\renewcommand{\phi}{\varphi}

\begin{document}

\begin{frontmatter}

\title{Application of large deviation theory to the mean-field $\varphi^4$-model}
\collab{Ingo Hahn \and Michael Kastner}
\address{Physikalisches Institut, Lehrstuhl f\"ur Theoretische Physik I, Universit\"at Bayreuth, 95440 Bayreuth, Germany}

\begin{abstract}
A large deviation technique is used to calculate the microcanonical entropy function $s(v,m)$ of the mean-field $\varphi^4$-model as a function of the potential energy $v$ and the magnetization $m$. As in the canonical ensemble, a continuous phase transition is found. An analytical expression is obtained for the critical energy $v_c(J)$ as a function of the coupling parameter $J$.
\end{abstract}

\end{frontmatter}

\section{Introduction}
Statistical physics is the branch of theoretical physics which provides a microscopic basis for the macroscopic theory of thermodynamics. Equilibrium statistical physics makes use of statistical ensembles, like the microcanonical or the canonical one. The basic thermodynamic function in the microcanonical ensemble is the entropy $s(v,m,\dotsc)$ as a function of some energy $v$, magnetization $m$, and/or further arguments, whereas in the canonical ensemble this role is played by the canonical free energy $f(T,h,\dotsc)$ as a function of the temperature $T$, a magnetic field $h$, and/or further arguments. The computation of thermodynamic functions from statistical mechanics is a formidable, in most cases impossible task in general, but, as a rule of thumb, it is more difficult in the microcanonical ensemble. This is the reason why, although the microcanonical ensemble is the most fundamental one of the statistical ensembles, one often resorts to the calculation of the canonical free energy $f$ instead. This is a reasonable way to proceed, as long as both options yield equivalent results. If this is not the case, one speaks of ensemble nonequivalence, and the occurrence of non-concave microcanonical entropy functions in systems with long-range interactions is a hallmark of this situation. Then, unlike in standard thermodynamics where the concavity of the entropy is assumed, the entropy $s$ cannot be obtained from the free energy $f$ by means of a Legendre transform.

Only in recent years the community of statistical physicists became aware that large deviation theory, a branch of probability theory, is a powerful tool to compute, among others, microcanonical quantities, at least for some classes of systems (see \cite{Ellis} for a mathematical treatise on large deviation theory with applications to statistical mechanics, or \cite{Touchette} for a less formal presentation of the subject). Explicit calculations of microcanonical quantities by means of large deviation theory can be found in \cite{BaBoDaRu} for some models of physical interest (the infinite range Potts model, the Hamiltonian mean-field model, and the Colson-Bonifacio model of a free electron laser).

We start by briefly reviewing some results of large deviation theory and their relation to statistical mechanics in Sec.~\ref{sec:ldt}. In Sec.~\ref{sec:model}, the microcanonical entropy of the $\varphi^4$-model without interaction is calculated using large deviation theory. From this result, the entropy of the mean-field $\varphi^4$-model in the presence of interactions is deduced. The physical implications of the result are discussed, followed by some conclusions in Sec.~\ref{sec:concl}.  

\section{Large deviation theory, a tool in statistical mechanics}
\label{sec:ldt}
This section contains a brief, informal review of some statements of large deviation theory and how they can be applied to statistical mechanics. Consider a sequence of independent, identically distributed (i.i.d.) random variables $X_i\in\RR^d$ with probability distribution $p(X)$, the empirical mean $\mathfrak{S}_N=\frac{X_1+\dots+X_N}{N}$, and the mean (or expectation) value $\mu:=\EE[X]:=\int\,\dd X\,X\,p(X)$. The law of large numbers tells us that, in the limit of large $N$, the probability $\PP$ of an event with $\mathfrak{S}_N\ne\mu$ converges to zero. The form of this convergence is the central issue of large deviation theory, thus allowing predictions about very unlikely events. Under some prerequisites on the random variables $X_i$, Cram\'er \cite{Cramer} found that, in the limit of large $N$, the convergence is of exponentially decaying form in $N$,
\begin{equation}\label{eq:LDP}
-I(x)=\lim_{N\to\infty}\frac{1}{N}\ln\PP(\mathfrak{S}_N\in [x,x+\dd x]).
\end{equation}
The so-called rate function $I(x)$ depends on the value $x$ the empirical mean has to meet. Comparing \eqref{eq:LDP} to the definition of the Boltzmann entropy,
\begin{equation}
s(x)=\lim_{N\to\infty}\frac{1}{N}\ln \int_{\RR^{Nn}}\dd X^N\, \delta\left[x-\mathfrak{S}_N\left(X^N\right)\right],
\end{equation}
we see that $s(x)$ is identical to $-I(x)$ up to a physically irrelevant additive constant. Under suitable conditions on $X_i$, large deviation theory gives an instruction how to derive the rate function (and therefore the microcanonical entropy):
\begin{equation}\label{eq:entropiex}
s(x)=-(t_x,x)+\ln\EE[\ee^{(t_x,x)}],
\end{equation}
where $(,)$ denotes the usual scalar product and $t_x(x)$ is obtained by solving the conditional equation
\begin{equation}\label{eq:legendremaximum}
x\EE[\ee^{(t_x,X)}] =\EE\left[X\,\ee^{(t_x,X)}\right].
\end{equation}

\section{Application to the mean-field $\boldsymbol{\varphi^4}$-model}\label{sec:model}
The mean-field $\varphi^4$-model we want to study is characterized by a potential energy function of the form
\begin{equation}\label{eq:pot}
V(\varphi)=-\frac{J}{2 N}\left(\sum_{i=1}^N\varphi_i\right)^2+\sum_{i=1}^N \left(\tfrac{1}{4}\,\phi_i^4-\tfrac{1}{2}\,\phi_i^2\right),
\end{equation} 
where $N$ is the number of particles, $J$ is a coupling constant, and $\varphi=(\varphi_1,\cdots,\varphi_N)$ denotes the position in configuration space. Our aim is to compute the microcanonical configurational entropy of this model.

\subsection{The $\varphi^4$-model without interaction}
\label{sec:no_int}
First, we look at the $\varphi^4$-model without interaction, where $N$ particles move independently in an on-site potential of the form $\tfrac{1}{4}\,\phi_i^4-\tfrac{1}{2}\,\phi_i^2$. The potential energy is
\begin{equation}\label{eq:onsite}
V_{os}(\varphi)=\sum_{i=1}^N \left(\tfrac{1}{4}\,\phi_i^4-\tfrac{1}{2}\,\phi_i^2\right).
\end{equation} 

The first step in calculating the configurational entropy is to translate our problem into the language of the previous section. Considering the coordinates $\varphi_i\in\RR$ as random variables with flat distribution on some finite interval,
\begin{equation}
p(\varphi)=
\begin{cases}
\frac{1}{2\varphi_c} & \text{for $\varphi\in[-\varphi_c,\varphi_c]$},\\
0 & \text{else},
\end{cases}
\end{equation}
other random variables
\begin{equation}\label{eq:Xi}
X_i=\left(\frac{1}{4}\,\phi_i^4-\frac{1}{2}\,\phi_i^2,\varphi_i\right)\in\RR^2
\end{equation}
with potential energy and displacement of the $i$-th particle as components can be constructed from the $\varphi_i$. This form of \eqref{eq:Xi} is motivated by the fact that it allows to express the potential energy function \eqref{eq:pot} by the empirical mean $\mathfrak{S}_N(X)=(z_N,m_N)$, where
\begin{align}
z_N(\varphi)&=\frac{1}{N} \sum_{i=1}^N \frac{1}{4}\,\phi_i^4-\frac{1}{2}\,\phi_i^2, \\
m_N(\varphi)&=\frac{1}{N} \sum_{i=1}^N \varphi_i.
\end{align}
The associated macroscopical variables $x=(z,m)$ are the mean potential energy per particle $z$ and the mean magnetization (or displacement) per particle $m$.

Then, following Eqs.~\eqref{eq:entropiex} and \eqref{eq:legendremaximum}, the entropy of the non-interacting $\varphi^4$-model can be written as
\begin{equation}\label{entropietilde}
\tilde{s}(z,m)=-t_z z-t_m m
+\ln\int_{-\varphi_c}^{+\varphi_c} \dd\varphi\, \ee^{t_m \varphi + t_z \left(\frac{1}{4}\varphi^4 - \frac{1}{2}\varphi^2\right)},
\end{equation}
where $t_z=t_z(z,m)$ and $t_m=t_m(z,m)$ are obtained by solving the conditional equations
\begin{equation}\label{tvtmcondition}
\begin{split}
z&=\tfrac{1}{4}\,\omega_4(t_z,t_m)-\tfrac{1}{2}\,\omega_2(t_z,t_m),\\
m&=\omega_1(t_z,t_m),
\end{split}
\end{equation}
containing the integrals
\begin{equation}\label{eq:omega_k}
\omega_k(t_z,t_m)=\frac{\int_{-\varphi_c}^{+\varphi_c} \dd\varphi\, \varphi^k\,\ee^{t_m \varphi + t_z \left(\frac{1}{4}\varphi^4 - \frac{1}{2}\varphi^2\right)}}{\int_{-\varphi_c}^{+\varphi_c} \dd\varphi\, \ee^{t_m \varphi + t_z \left(\frac{1}{4}\varphi^4 - \frac{1}{2}\varphi^2\right)}} .
\end{equation}
We let $\varphi_c$ tend to infinity in the following.

\begin{figure}[ht]
\begin{center}
\epsfig{file=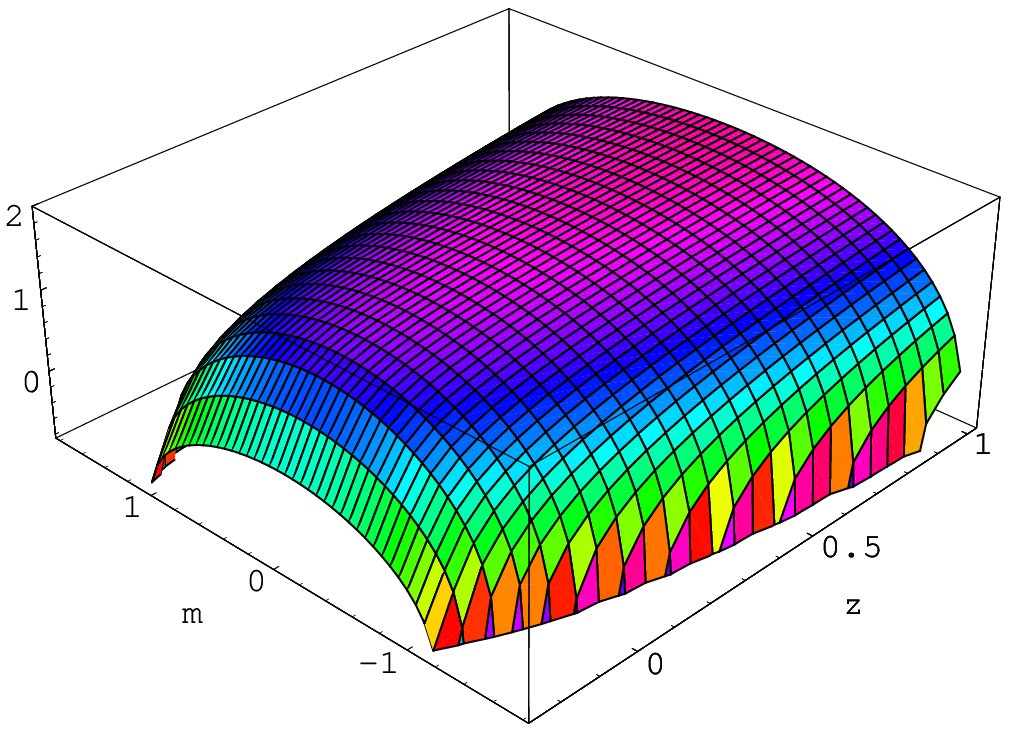,width=7cm}
\hspace{5mm}
\epsfig{file=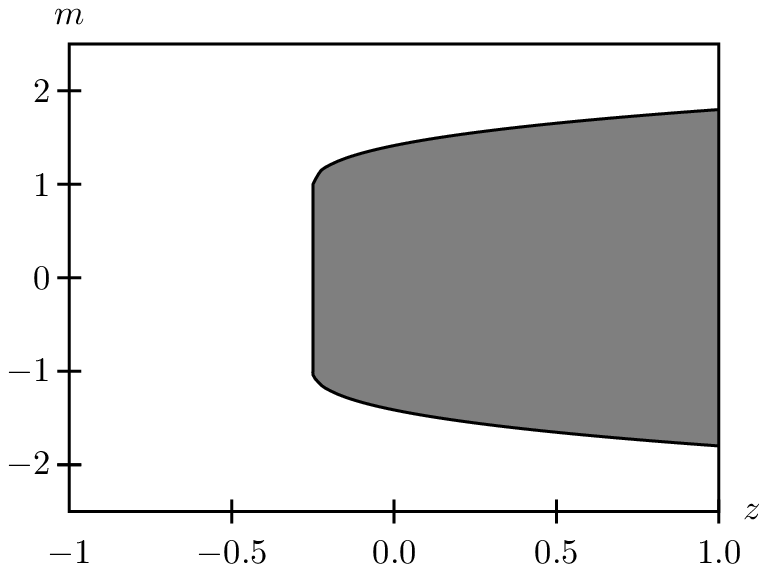,width=6cm}
\end{center}
\caption{\label{figuresmz}
Entropy $\tilde{s}(z,m)$ (left plot) and its domain (right plot, gray hatched area) of the $\varphi^4$-model without interaction from a numerical evaluation of \eqref{entropietilde}.}
\end{figure}

Solving \eqref{tvtmcondition} for $t_z$ and $t_m$ seems not feasible analytically. However, besides a numerical solution (see Fig. \ref{figuresmz}), some general properties of $\tilde{s}(z,m)$ may be of interest. From large deviation theory we know that $\tilde{s}(z,m)$ is a concave, infinitely many times differentiable function. Its domain can be determined by the following argument: The energy $V(\varphi_i)=\tfrac{1}{4}\,\phi_i^4-\tfrac{1}{2}\,\phi_i^2$ of any single particle is not bounded above, but it is bounded below by $-\tfrac{1}{4}$. Therefore also the mean potential energy $z$ is restricted to the interval $[-\tfrac{1}{4},\infty[$. For any value of $z\geqslant\tfrac{1}{4}$, there exists a maximal and a minimal value of the magnetization $m_{max}(z)=-m_{min}(z)$ which is attained when $\varphi_i=\varphi_j$ for all $i,j$. So $m_{max/min}(z)$ is determined by $z=\tfrac{1}{4}\,m_{max/min}^4-\tfrac{1}{2}\,m_{max/min}^2$. Starting from a microstate with $(z,m_{max})$, we can always find another microstate with $m\in[-m_{max},m_{max}]$ by changing the sign of an arbitrary fraction of the values of the $\phi_i$. For an illustration of the domain of $\tilde{s}(z,m)$ see Fig.~\ref{figuresmz}.
For any fixed value of $z\geqslant\frac{1}{4}$, the maximum of $\tilde{s}$ with respect to $m$ is located at zero magnetization.

\subsection{The $\varphi^4$-model with mean-field interaction}
\begin{figure}[ht]
\begin{center}
\epsfig{file=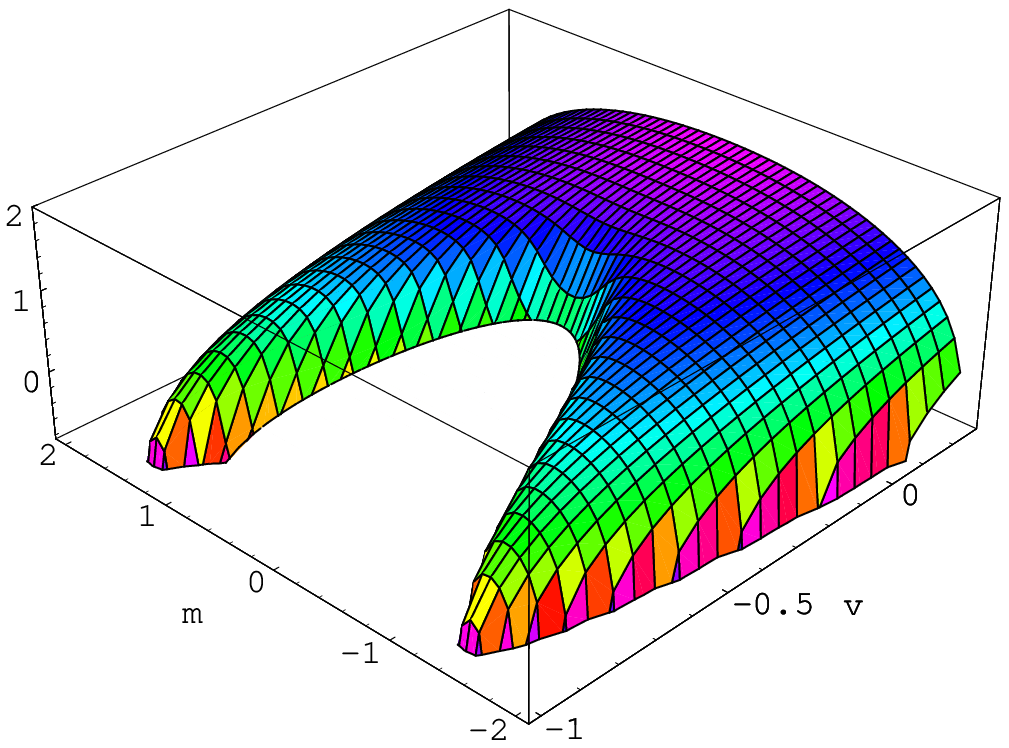,width=7cm}
\hspace{5mm}
\epsfig{file=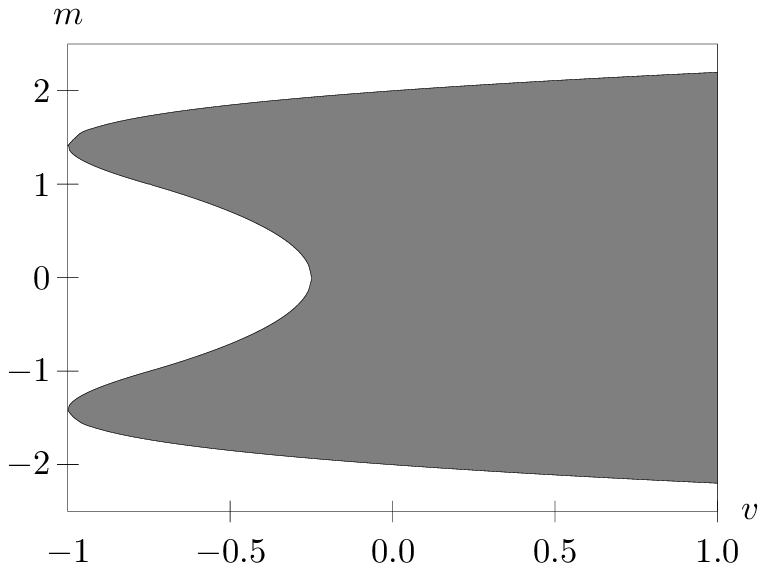,width=6cm}
\end{center}
\caption{\label{figuresmv}
Entropy $s(v,m)$ of the mean-field $\varphi^4$-model for coupling constant $J=1$ (left plot) and its domain (right plot, gray hatched area). $s(v,m)$ is obtained by a deformation (variable transformation) of the entropy $\tilde{s}(z,m)$ of the non-interacting model (see Fig.~\ref{figuresmz}). Below a critical value $v_c$ of the potential energy $v$, the maximum of $s$ with respect to the magnetization $m$ is located at a non-zero value of $m$.}
\end{figure}

Turning to the study of the interacting model, we can now profit from our particular choice of the random variables $X_i$ in Eq.~\eqref{eq:Xi}, resulting in the macroscopic variables $(z,m)$. From the form \eqref{eq:pot} of the potential energy function $V$, the mean potential energy per particle can be written as
\begin{equation}
v_N(\varphi)=z_N(\varphi)-\tfrac{J}{2}\,m_N(\varphi)^2,
\end{equation}
and for the corresponding macroscopic variable $v$ the equality
\begin{equation}
v=z-\tfrac{J}{2}\,m^2
\end{equation}
holds. As a consequence, the entropy $s(v,m)$ of the mean-field $\varphi^4$-model is obtained from the entropy of the non-interacting model by a simple transformation of variables,
\begin{equation}
s(v,m)=\tilde{s}(v+\tfrac{J}{2}\,m^2,m).
\end{equation}
A plot of the resulting function is shown in Fig.~\ref{figuresmv}.
For large enough fixed values of the potential energy $v$, the maximum of $s$ with respect to $m$ is again located at zero magnetization. Below a critical value $v_c$ of the potential energy, however, this is not the case anymore, indicating the occurrence of a phase transition. The transition is found to be a continuous one, and $v_c$ is determined by a change of curvature of $s$ with respect to $m$ along the line of zero magnetization,
\begin{equation}
\frac{\partial^2 s(v_c,m)}{\partial m^2}\bigg|_{m=0}=0.
\end{equation}
From the results of Sec.~\ref{sec:no_int}, the critical potential energy $v_c$ in dependence of the coupling constant $J$ can be written as
\begin{equation}\label{criticalenergy}
v_c(J)=\frac{\tfrac{1}{J}-1}{4\,t_{z_c}(J)},
\end{equation}
where $t_{z_c}$ is defined implicitly by
\begin{equation}\label{eq:detcrittz}
J\,t_{z_c}\,\omega_2(t_{z_c},0)=-1.
\end{equation}
It follows immediately that $v_c(1)=0$ and, since $t_z<0$, that $v_c(J)>0$ for all $J>0$. The integral $\omega_2(t_{z_c},0)$ in \eqref{eq:detcrittz} can be rewritten in terms of modified Bessel functions of the first kind $I_k$, yielding
\begin{equation}\label{besselintegral}
\omega_2(t_z,0)=\frac{1}{2}\left(1+\frac{I_{-\frac{3}{4}}(-\frac{t_z}{8})+I_{\frac{3}{4}}(-\frac{t_z}{8})}{I_{-\frac{1}{4}}(-\frac{t_z}{8})+I_{\frac{1}{4}}(-\frac{t_z}{8})}\right).
\end{equation}
This allows to derive asymptotic expansions of $v_c(J)$ for small and large positive $J$, respectively, from the known expansions of these Bessel functions.

\paragraph*{Small $J>0$:} Inserting the asymptotic expansion of \eqref{besselintegral} for large negative $t_{z_c}$ into \eqref{eq:detcrittz}, solving for $t_{z_c}(J)$ and substituting into \eqref{criticalenergy} finally yields 
\begin{equation}\label{eq:vc_smallJ}
v_c(J)=-\tfrac{1}{4}+\tfrac{1}{2}\,J+{\mathcal{O}}(J^2).
\end{equation}

\paragraph*{Large $J>0$:} Analogously, an expansion of \eqref{eq:detcrittz} for small negative $t_{z_c}$ yields an expression for the critical energy in the limit of large $J$,
\begin{equation}\label{eq:vc_largeJ}
v_c(J)=a^2 J^2-\left(2\,a^2-\tfrac{1}{4}\right)J+\left(\tfrac{5\,a^2}{4}-\tfrac{3}{8}+\tfrac{1}{64 \,a^2}\right)+{\mathcal{O}}\left(\tfrac{1}{J}\right),
\end{equation}
with $a=\Gamma(\frac{3}{4})/\Gamma(\frac{1}{4})\approx 0.338$.

\begin{figure}[ht]
\begin{center}
\epsfig{file=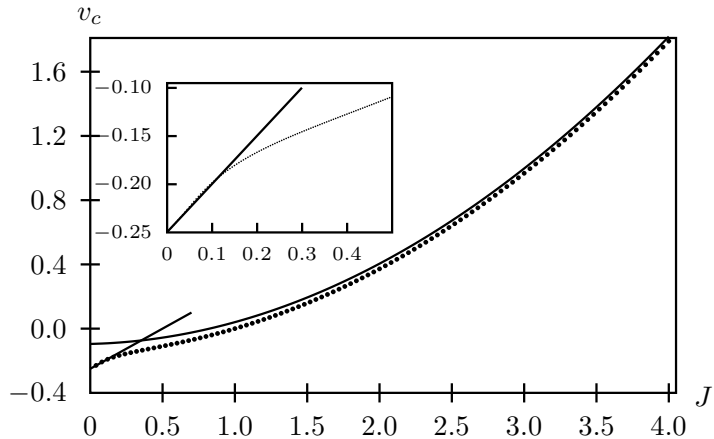,width=9.5cm}
\end{center}
\caption{\label{figurecritical}
Critical energy $v_c$ as a function of the coupling constant $J$ from numerical computation (dots), confronted with the two asymptotic expansions (\ref{eq:vc_smallJ}) and  (\ref{eq:vc_largeJ}).
}
\end{figure}

Together with the numerical result from the evaluation of Eqs.~\eqref{criticalenergy} and \eqref{eq:detcrittz}, these two asymptotic expansions are plotted in Fig.~\ref{figurecritical}. 

\section{Conclusions}
\label{sec:concl}
The microcanonical entropy $s(v,m)$ as a function of the potential energy $v$ and the magnetization $m$ of the mean-field $\phi^4$-model has been calculated using a large deviation technique. For such long-range interactions, and in contrast to standard thermodynamics, the entropy need not be a concave function. In fact, for a ferromagnetic coupling $J>0$, a non-concavity in $s$ is found, and therefore, when comparing this result to the canonical free energy $f(T,h)$ as a function of the temperature $T$ and the magnetic field $h$, nonequivalence of ensembles is observed. For any fixed value of $v$ above a critical energy $v_c$, the maximum of $s(v,m)$ with respect to $m$ is attained at $m=0$, whereas below $v_c$ it is attained at $m\ne 0$, corresponding to a continuous order phase transition at $v_c$. An exact, implicit expression for the critical energy $v_c(J)$ as well as expansions for small and large positive $J$ are derived.

Maximizing $s(v,m)$ over $m$ gives a concave entropy function
\begin{equation}
\hat{s}(v)=\max_{m}\left[s(v,m)\right],
\end{equation}
showing ensemble equivalence when compared to the canonical free energy $\hat{f}(T)=\max_{h}[f(T,h)]$.  In fact, the critical energy $v_c(J)$ from our microcanonical calculation corresponds to the critical temperature $T_c(J)$ of the canonical result reported in \cite{AnAnRuZa:04}.

\end{document}